\documentclass[aps,pra,showpacs,amsmath,lengthcheck,superscriptaddress]{revtex4-2}

\usepackage{graphicx}\graphicspath{{figures/}}
\usepackage[caption=false]{subfig}
\usepackage[colorlinks=true,citecolor=blue,linkcolor=blue,urlcolor=blue]{hyperref}%

\renewcommand{\d}{\partial}
\newcommand{\bra}[1]{\langle #1|}
\newcommand{\ket}[1]{|#1\rangle}
\newcommand{\braket}[2]{\langle #1 | #2 \rangle}

\newcommand{\tr}[1]{\mbox{tr}\left\{ #1 \right\}}
\newcommand{\re}[1]{\mbox{Re}\left\{ #1 \right\}}
\newcommand{\bla}{\noindent bla\\bla\\bla\\bla\\bla}

\begin{document}

\title{The three phases of quantum annealing: fast, slow, and very slow}

\author{Artur Soriani}
\email{asorianialves@gmail.com}
\affiliation{Instituto de F\'isica `Gleb Wataghin', Universidade Estadual de Campinas, 13083-859, Campinas, S\~{a}o Paulo, Brazil}
\author{Pierre Naz\'e}
\email{p.naze@ifi.unicamp.br}
\affiliation{Instituto de F\'isica `Gleb Wataghin', Universidade Estadual de Campinas, 13083-859, Campinas, S\~{a}o Paulo, Brazil}
\author{Marcus V. S. Bonan\c{c}a}
\affiliation{Instituto de F\'isica `Gleb Wataghin', Universidade Estadual de Campinas, 13083-859, Campinas, S\~{a}o Paulo, Brazil}
\author{Bart{\l}omiej Gardas}
\affiliation{Institute of Theoretical and Applied Informatics, Polish Academy of Sciences, Ba{\l}tycka 5, 44-100 Gliwice, Poland}
\author{Sebastian Deffner}
\affiliation{Department of Physics, University of Maryland, Baltimore County, Baltimore, Maryland 21250, USA}
\affiliation{Instituto de F\'isica `Gleb Wataghin', Universidade Estadual de Campinas, 13083-859, Campinas, S\~{a}o Paulo, Brazil}

\date{\today}

\begin{abstract}
Currently, existing quantum annealers have proven themselves as viable technology for the first practical applications in the noisy-intermediate-scale-quantum era.
However, to fully exploit their capabilities, a comprehensive characterization of their finite-time excitations is instrumental.
To this end, we develop a phase diagram for driven Ising chains, from which the scaling behavior of the excess work can be read off as a function of process duration and system size.
``Fast'' processes are well described by the Kibble-Zurek mechanism; ``slow'' processes are governed by effective Landau-Zener dynamics; and ``very slow'' processes can be approximated with adiabatic perturbation theory.
\end{abstract}


\maketitle

\section{\label{sec:Intro}Introduction}

It has been four decades since Feynman first proposed to harness genuine quantum properties to build better, more powerful computers~\cite{Feynman1982,preskill2021}.
However, only now do we finally appear to be standing at the beginning of the \emph{quantum information age}~\cite{fuchs2011}, which is evidenced by national as well as international quantum initiatives \cite{Riedel2019QST,Yamamoto2019QST,Sussman2019QST,Roberson2019QST} and the first demonstrations of verifiable quantum advantage~\cite{Arute2019Nature,Zhong2020}.
Yet it may take a little while longer before the first practically useful and fault-tolerant quantum computers become widely available~\cite{Sanders2017}.
In the meanwhile, so-called \emph{noisy intermediate-scale quantum} (NISQ) may already be useful for special applications~\cite{Preskill2018quantum}.

For instance, it was shown only very recently that already current generations of the D-Wave machine can handle complex, realistic problems in quantum simulation \cite{King2018,Harris2018,King2022} and in classical optimization \cite{Yarkoni2021}, such as conflict management in existing railway networks~\cite{domino2021trains}, although \emph{quantum advantage} has not been reached yet in this context.
As a quantum annealer, solving problems with the D-Wave machine relies on adiabatic quantum computing~\cite{nielsen2010}, at least in an ideal situation.
However, like all real systems, the D-Wave machine is subject to noise~\cite{Gardas2018SR,Gardas2019PRA}.
And if this system is ever going to be implemented as a computer for real-life applications, complete characterization is instrumental.
To this end, the scaling properties of the nonadiabatic excitations have been thoroughly investigated~\cite{Bartek2018,Bando2020}.
Despite significant deviations from the expected behavior (due to environmental noise), the D-Wave chip seems to, indeed, implement a quantum Ising model in the transverse field~\cite{Bando2020}.

However, even the ideal case of an isolated, driven quantum Ising model is far from trivial to fully analyze.
Typically, the dynamics has to be solved numerically~\cite{Francuz2016PRB}, and approximate, less computationally intensive approaches appear to be highly desirable.
It has been well established that for “fast” (but not too fast) processes the dynamics is well described by the Kibble-Zurek mechanism~\cite{Zurek2005,Francuz2016PRB,Puebla2020PRR}, whereas for “slow” (but not too slow) driving the Landau-Zener formula becomes applicable~\cite{Dziarmaga2005}.

In the present work, we give a comprehensive characterization of the dynamical properties of the driven quantum Ising chain in the transverse field.
To this end, we show that for ``very slow'' processes the Landau-Zener formula becomes inapplicable, and rather adiabatic perturbation theory (APT)~\cite{Messiah1962} properly describes the dynamics.
Moreover, we make the distinction between fast, slow, and very slow regimes rigorous by determining the crossover points between the three different regimes.
As a main result, we obtain a \emph{dynamic phase diagram} (in contrast to the usual equilibrium phase diagrams) for the predicted dynamical properties as a function of the number of Ising spins and the duration of the driving.

The present analysis seeks to be as self-contained as possible.
Thus, we briefly outline adiabatic perturbation theory in Sec.~\ref{sec:Definitions}, before we work through a pedagogical example, namely the Landau-Zener model in Sec.~\ref{sec:LandauZener}.
A complete analysis of the time-dependent quantum Ising model is discussed in Sec.~\ref{sec:IsingChain}, whose experimental consequences for the D-Wave machine are elaborated in Sec.~\ref{sec:DWave}.
The analysis is concluded in Sec.~\ref{sec:Conclusion}.

\section{\label{sec:Definitions}Preliminaries}

We start by establishing notions and notations, with a brief review of adiabatic perturbation theory.

\paragraph*{Quantum excess work.}

In the present analysis, we focus on ideal quantum annealing and thus consider only isolated quantum systems.
We write the Hamiltonian as $H(\lambda)=\sum_n E(\lambda) \ket{n(\lambda)}\bra{n(\lambda)}$, where $\lambda$ is a time-dependent, external control parameter $\lambda = \lambda(t)$, which is varied for a duration $\tau=t_f-t_i$ such that $\lambda(t_i) = \lambda_i$ to $\lambda(t_f) = \lambda_f$.

As usual in quantum annealing, we assume that initially,  the system is prepared in its ground state $\ket{\psi(t_i)}=\ket{g(\lambda_i)}$, and the dynamics is given by the Schr\"odinger equation $i \ket{\dot{\psi}(t)} = H(\lambda) \ket{\psi(t)}$, where we set $\hbar=1$ and the dot denotes the derivative with respect to time. 

For such scenarios~\cite{Deffner2017PRE,Deffner2019book}, the \emph{excess work} is defined as total variation of the average energy minus the difference in initial and final ground-state energies.
Hence, we can write
\begin{equation} \label{eq:ExcessWork}
W_\mathrm{ex} = \sum_{n \neq g} p_n \left[ E_n(\lambda_f) - E_g(\lambda_f) \right],
\end{equation}
where $p_n$ is the unitary transition probability, $p_n = \left|\braket{n(\lambda_f)}{\psi_g(t_f)} \right|^2$.
Further, $\ket{\psi_g(t_f)}$ is the initial ground state $\ket{g(\lambda_i)}$, evolved under the time-dependent Schr\"odinger equation.

In the following, we will analyze the scaling properties of the excess work $W_\mathrm{ex}$ for systems that cross a quantum critical point (QCP).
For such scenarios it has been demonstrated that $W_\mathrm{ex}$ fully characterizes the phase transition~\cite{Silva2008PRL,Mascarenhas2014PRE,Fusco2014PRX,Campbell2016PRB} and that it even exhibits Kibble-Zurek scaling~\cite{Francuz2016PRB}.
However, for general systems, fully analyzing the dynamical properties is a computationally hard problem, which is why sudden quenches are often considered~\cite{Silva2008PRL,Mascarenhas2014PRE,Fusco2014PRX,Campbell2016PRB}.
In contrast, here we develop approximate methods that allow us to determine $W_\mathrm{ex}$ for any duration of the process $\tau$ (relevant to quantum annealing), namely, fast, slow, and very slow.

\paragraph*{\label{sec:APT}Adiabatic perturbation theory}

To complement existing, approximate methods, we employ adiabatic perturbation theory~\cite{Messiah1962,Rigolin2008,Morita2008}.
This approach provides corrections to the adiabatic solution in powers of $1/\tau$. Hence, APT is a perturbation theory for very slow processes. 

For our purposes, that is, for systems initially prepared in the ground state, we can write
\begin{equation} \label{eq:APTEvolution}
\ket{\psi_g(t)} = \exp[i\phi_g(t)]\, \sum_{p=0}^{\infty} \ket{\psi_g^{(p)}(t)},
\end{equation}
where
\begin{equation} \label{eq:APTExpansion}
\ket{\psi_g^{(p)}(t)} = \sum_m C_{m}^{(p)}(t)\, \ket{m(\lambda)},
\end{equation}
is the $p$-th order correction written in the basis of instantaneous eigenstates of $H(\lambda)$.
As always,
\begin{equation} \label{eq:AdiabaticPhases}
\phi_n(t) = - \int_{t_i}^t E_n\textbf{(}\lambda(t')\textbf{)} dt'+ i \int_{t_i}^t \braket{n\textbf{(}\lambda(t')\textbf{)}}{\dot{n}\textbf{(}\lambda(t')\textbf{)}} dt'\,.
\end{equation}
From Eqs.~\eqref{eq:APTEvolution}--\eqref{eq:AdiabaticPhases} the transition probability $p_n$ can be computed to arbitrary order.

The coefficients $C_{m}^{(p)}(t)$, for $p>0$, can be systematically calculated.
For example, the expression for $p=1$ and $m \neq g$ reads
\begin{equation} \label{eq:APTFirstOrderCoefficient}
C_{m}^{(1)}(t) = i \left( \frac{M_{mg}(t)}{E_{mg}(\lambda)} -  \frac{M_{mg}(t_i)}{E_{mg}(\lambda_i)}\,\exp[i \phi_{mg}(t)] \right),
\end{equation}
where $E_{mn}(\lambda) = E_m(\lambda) - E_n(\lambda)$, $\phi_{mn}(t) = \phi_m(t) - \phi_n(t)$, and $M_{mn}(t)$ is given by
\begin{equation}
\label{eq:APTMatrix}
M_{mn}(t) = \braket{m(\lambda)}{\dot{n}(\lambda)}= - \dot{\lambda}(t) \frac{\bra{m(\lambda)} \d_{\lambda} H(\lambda) \ket{n(\lambda)}}{E_{mn}(\lambda)}\,,
\end{equation}
where the second equality is valid only for $m \neq n$.

In the following, we will consider only driving protocols with fixed $\lambda_i$ and $\lambda_f$.
Therefore, $\dot{\lambda} \propto \tau^{-1}$, which determines the magnitude of $C_{m}^{(1)}(t)$ in Eq.~\eqref{eq:APTFirstOrderCoefficient}.
Similarly, $C_{m}^{(2)}(t)$ contains $\ddot{\lambda}$ and $\dot{\lambda}^2$, both of which are proportional to $\tau^{-2}$, with analogous notation for higher orders.
Hence, for $\tau \to \infty$, all terms but the first in Eq.~\eqref{eq:APTEvolution} vanish, and we recover the adiabatic limit.

The range of validity of APT is governed by~\cite{Rigolin2008}
\begin{equation} \label{eq:APTValidityCondition}
\left\vert \frac{M_{mn}(t)}{E_{mn}(\lambda)} \right\vert \ll 1\,,
\end{equation}
which is not met when $E_{mn}(\lambda)$ is small in comparison to $\dot{\lambda}(t)$ at any point of the process.
Thus, we would expect a breakdown of the approximation for processes that rapidly cross a QCP.

\section{\label{sec:LandauZener}Generalized Landau-Zener model}

To demonstrate the utility of APT and where it fits in comparison to other approximate techniques, we treat a simple, pedagogical example first --- the Landau-Zener (LZ) model \cite{Landau1932,Zener1932,Stuckelberg1932,Majorana1932} for arbitrary driving. Namely,
\begin{equation} \label{eq:LZHamiltonian}
H_{LZ}(\lambda) = \Delta\, \lambda \sigma^z + J \sigma^x,
\end{equation}
where $\Delta$ and $J$ are positive constants and $\sigma^z$ and $\sigma^x$ are Pauli matrices.
Note that the avoided crossing is the simplest representation of a QCP, and the LZ model even exhibits a scaling reminiscent of the Kibble-Zurek mechanism \cite{Damski2005}.

Defining the eigenstates of $ \sigma^z$ as $\sigma^z \ket{\downarrow^z} = - \ket{\downarrow^z} $ and $\sigma^z \ket{\uparrow^z} = \ket{\uparrow^z} $, the energy eigenstates become
\begin{equation} \label{eq:LZEigenStates}
\begin{array}{c}
\ket{-(\lambda)} = \cos \theta(\lambda) \ket{\downarrow^z} - \sin \theta(\lambda) \ket{\uparrow^z}, \\
\ket{+(\lambda)} = \sin \theta(\lambda) \ket{\downarrow^z} + \cos \theta(\lambda) \ket{\uparrow^z},
\end{array}
\end{equation}
where
\begin{equation} \label{eq:LZTheta}
\theta(\lambda) = \frac{1}{2} \arctan\left(J/\Delta \lambda\right),
\end{equation}
and the eigenvalues are
\begin{equation} \label{eq:LZEigenEnergies}
E_{\pm}(\lambda) = \pm E(\lambda) = \pm \sqrt{\Delta^2 \lambda^2 + J^2}.
\end{equation}
Note that $g = -1$ corresponds to the ground state.

\begin{figure}
\includegraphics[width=\columnwidth]{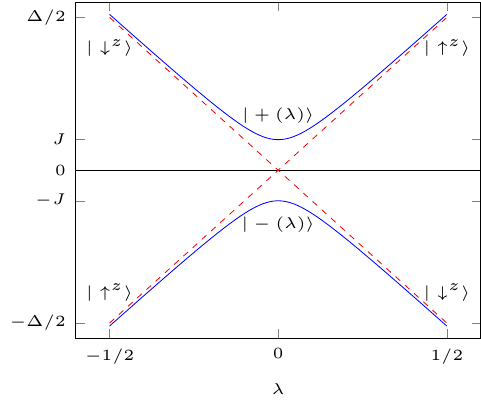}

\caption{\label{fig:LZeigenenergies}
Eigenenergies (blue solid lines) of the Landau-Zener model \eqref{eq:LZEigenEnergies} for $\Delta \gg J$ together with eigenvalues (red dashed lines) of $\Delta \lambda \sigma^z$.
}
\end{figure}

The gap $2E(\lambda)$ between eigenstates has a minimum for $\lambda = 0$, where it is equal to $2J$.
Figure~\ref{fig:LZeigenenergies} depicts the avoided crossing of the energy levels for $\lambda_f = 1/2 = -\lambda_i$ and for $\Delta \gg J$.
The dashed red lines represent the eigenvalues of the operator $\Delta \lambda \sigma^z$.
Observe that, at the end points of the process, the eigenstates of $\sigma^z$ and $H(\lambda)$ coincide approximately (apart from irrelevant change of signs).
However, there is a switch halfway through the process: at the beginning, we have $\ket{\downarrow^z} \approx \ket{+(\lambda_i)}$, while at the end, $\ket{\downarrow^z} \approx \ket{-(\lambda_f)}$.

The solution of Schrödinger's equation can be expressed as a linear combination of the eigenstates of $\sigma^z$,
\begin{equation} \label{eq:LZDynamicState}
\ket{\psi(t)} = u(t) \ket{\uparrow^z} + v(t) \ket{\downarrow^z},
\end{equation}
and we obtain
\begin{equation} \label{eq:LZDifferentialEquations}
\begin{split}
i\, \dot{u}(t) &=  \Delta \lambda(t) u(t) + J v(t), \\
i\, \dot{v}(t) &=  J u(t) - \Delta \lambda(t) v(t).
\end{split}
\end{equation}

It is interesting to point out that, in the original treatment of the LZ model \cite{Landau1932,Zener1932,Stuckelberg1932,Majorana1932}, only linear protocols of infinite duration were considered,
\begin{equation} \label{eq:LZOriginal}
\lambda(t) = t, \qquad -\infty < t < \infty.
\end{equation}
In this case, Eqs.~\eqref{eq:LZDifferentialEquations} can be solved analytically \cite{Zener1932,Vitanov1996}.
However, the exact solution is written as sums of parabolic cylinder functions with complex parameters and arguments, which make extracting their behavior computationally intensive.
Moreover, in the present work, we are interested in processes of finite duration $\tau$ that keep the initial and final values of $\lambda$ fixed, no matter the value of $\tau$. 

In any case, $W_\mathrm{ex}$ \eqref{eq:ExcessWork} can be expressed as
\begin{equation} \label{eq:LZExcessWork}
W_\mathrm{ex}(\tau) = 2 E(\lambda_f) p_+(\tau),
\end{equation}
where the transition probability  from the initial ground state to $\ket{+(\lambda_f)}$ now reads $p_+ = \left| \braket{+(\lambda_f)}{\psi_-(t_f)} \right|^2$.

For the sake of simplicity, we will continue the analysis with a linear protocol,
\begin{equation} \label{eq:LZLinearProtocol}
\lambda(t) = \frac{t}{\tau}, \qquad -\frac{\tau}{2} \leq t \leq \frac{\tau}{2}.
\end{equation}
It is worth emphasizing that, in contrast to the original LZ model~\cite{Landau1932,Zener1932,Stuckelberg1932,Majorana1932}, our protocol~\eqref{eq:LZLinearProtocol} obeys $\dot{\lambda}(t) \to 0$ as $\tau \to \infty$, whereas in the original treatment the rate $\dot{\lambda}(t)$ was held constant.

Thus, there is no immediate reason to believe that the Landau-Zener formula (LZF) is applicable.
Expressed in our notation, the LZF reads
\begin{equation} \label{eq:LZFormula}
p_+^\mathrm{LZ}(\tau) = \exp\left(- \pi\,J^2\tau/\Delta \right).
\end{equation}
Nevertheless, we will see that for specific values of $J$, $\Delta$, and $\tau$, Eq.~\eqref{eq:LZFormula} approximates the exact dynamics remarkably well.

On the other hand, the transition probability can also be computed from APT.
We have, from Eqs.~\eqref{eq:APTExpansion} and \eqref{eq:APTFirstOrderCoefficient} and the definition of $p_+$,
\begin{equation} \label{eq:LZ_APTTransitionProbability_Linear}
p_+^\mathrm{APT}(\tau) = \frac{1}{16} \left( \frac{\Delta}{J^2\tau} \right)^2 \left| \frac{J^3}{E^3(\lambda_f)} -  \frac{J^3 \exp[-2i\phi(\tau)]}{E^3(\lambda_i)} \right|^2\,.
\end{equation}
As before,  $\phi$ is the dynamic phase, which we can write as $\phi(\tau)=- \tau \int_{\lambda_i}^{\lambda_f} E(\lambda) d\lambda$.
Note that the dynamic phase governs the overall oscillatory behavior, which we will ``average out'' in the following analysis.
Finally, APT is expected to apply if condition~\eqref{eq:APTValidityCondition} is met throughout the entire process.
In the present case, this translates to $J^2 \tau/\Delta \gg 1$.

\begin{figure*}
\subfloat[\label{fig:LZWorkMainPlot}]{\includegraphics[width=.33\textwidth]{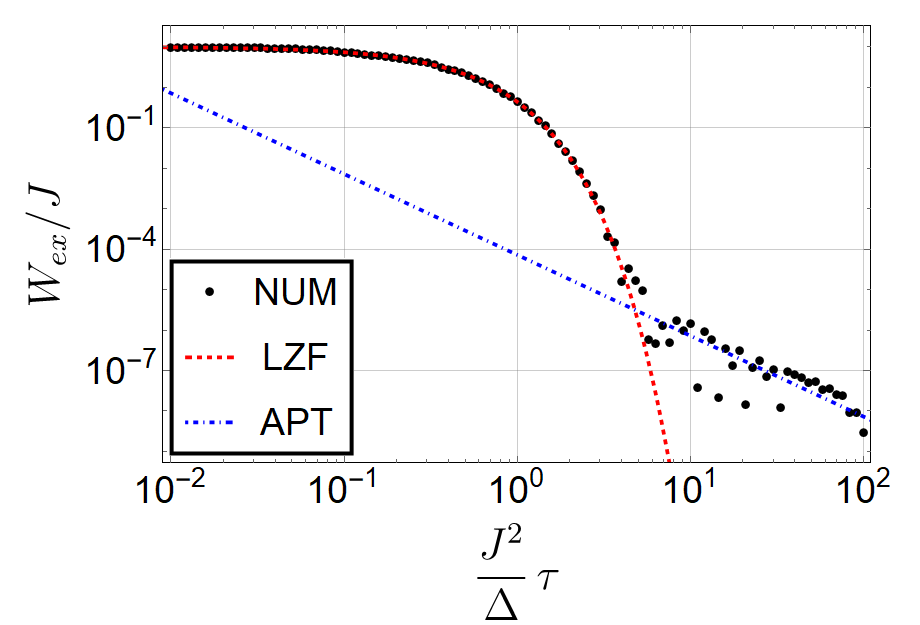}}
\subfloat[\label{fig:LZ_APTOscillations}]{\includegraphics[width=.33\textwidth]{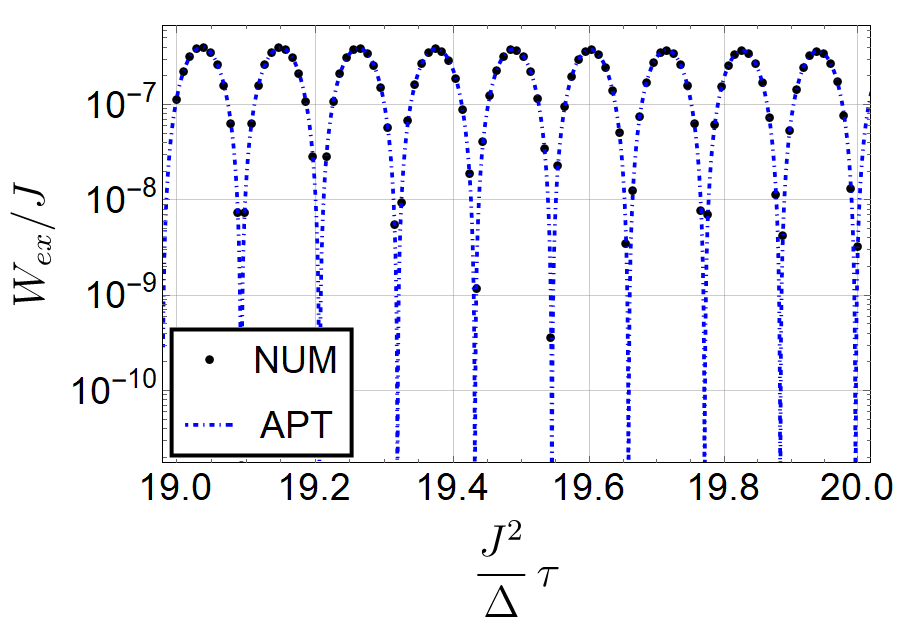}}
\subfloat[\label{fig:LZ_LZFormula}]{\includegraphics[width=.33\textwidth]{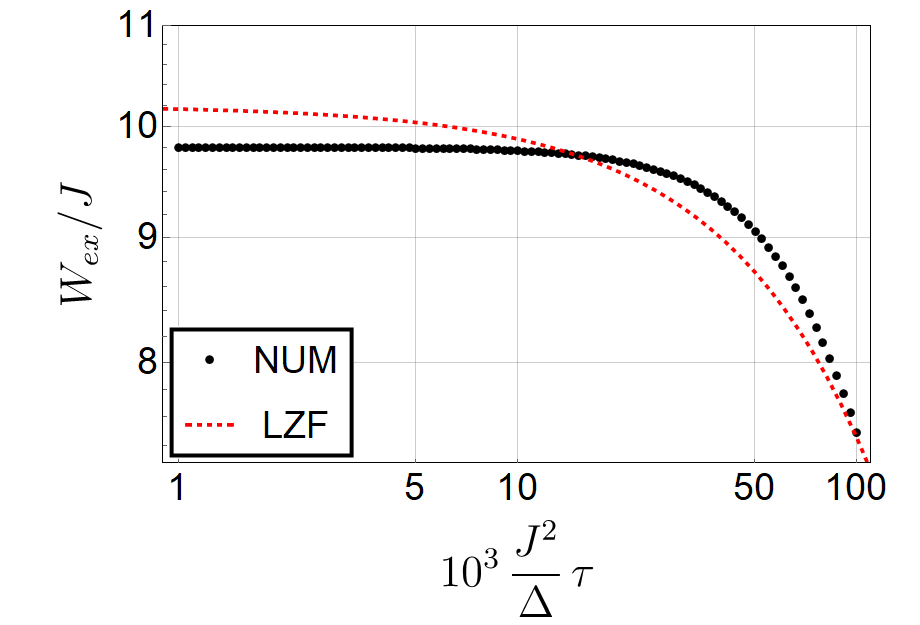}}
\caption{\label{fig:LZPlot}
Excess work~\eqref{eq:LZExcessWork} as a function of process duration for $\Delta/J = 10$.
Black dots represent the numerical solution, the red dashed line is computed from the LZF~\eqref{eq:LZFormula}, and the blue dash-dotted line is computed from APT~\eqref{eq:LZ_APTTransitionProbability_Linear}.
(a) The LZF-APT crossover, where the line corresponding to APT is phase-averaged.
(b) Zoom of a $\tau$ range where APT is valid, with oscillations included.
(c) Zoom of a $\tau$ range where LZF fails, as per Eq.~\eqref{eq:LZStrongCouplingValidity}.
}
\end{figure*}

In Fig.~\ref{fig:LZPlot} we compare the LZF~\eqref{eq:LZFormula} with the result of APT~\eqref{eq:LZ_APTTransitionProbability_Linear} and the numerically exact solution (from standard fourth-order Runge-Kutta).
On the $x$ axis, we have $J^2\tau/\Delta$, which allows us to unambiguously identify the range of validity of the approximate methods.
For ease of representation, the prediction from APT is ``phase averaged'' to remove the dynamical oscillations alluded to above.
Note that this fact is depicted in Fig.~\ref{fig:LZ_APTOscillations} for a small interval in $\tau$.

For $J^2\tau/\Delta < 1$, we observe striking agreement between the LZF formula~\eqref{eq:LZFormula} and the exact solution.
This fact was elucidated in Ref.~\cite{Vitanov1996}.
In that work, the authors considered finite-time driving of the LZ Hamiltonian~\eqref{eq:LZHamiltonian}, such that the initial and final eigenvalues diverge in the limit $\tau \to \infty$.
Our present situation can be mapped exactly to the dynamics considered in Ref.~\cite{Vitanov1996}, provided
\begin{equation} \label{eq:LZStrongCouplingValidity}
\frac{J^2\tau}{\Delta} + \left( \frac{\Delta}{2J} \right)^2 \frac{J^2\tau}{\Delta} \gg 1\,.
\end{equation}
Equation~\eqref{eq:LZStrongCouplingValidity} consists of two independently positive terms.
Hence, only one of the terms needs to be large for Eq.~\eqref{eq:LZStrongCouplingValidity} to hold.
In our case, we have $J^2\tau/\Delta \gg 1$ for APT to apply.
In the opposite limit, i.e., if $J^2\tau/\Delta$ is small, Eq.~\eqref{eq:LZStrongCouplingValidity} is governed by $\Delta/J \gg 1$.
In this limit, Ref.~\cite{Vitanov1996} demonstrated that (at least in leading order) the LZF is, in fact, a good approximation of the exact solution.
Further analysis of the validity of the LZF in finite time, including a nonanalytic APT approach, is given in Ref. \cite{DeGrandi2010}.

We can conclude that, for slow enough processes, the range of validity of the LZF formula~\eqref{eq:LZFormula} crosses over to APT~\eqref{eq:LZ_APTTransitionProbability_Linear}.
The crossover point $\tau_c$ is determined by
\begin{equation} \label{eq:LZCrossoverTimeDefinition}
p_+^\mathrm{LZ}(\tau_c) = p_+^\mathrm{APT}(\tau_c),
\end{equation}
with a $\tau$-independent phase $\phi$.
The exact solution of Eq.~\eqref{eq:LZCrossoverTimeDefinition} can be written as a function of Lambert's function $W_{-1}$~\cite{Lambert1996}.
For $\Delta/J \gg 1$, the asymptotic expression for $\tau_c$ becomes
\begin{multline} \label{eq:LZCrossoverTime}
\frac{J^2}{\Delta} \tau_c = \frac{2}{\pi} \Bigg\{ \ln \left[ \frac{4}{\pi} \left( \frac{\Delta}{2J} \right)^3 \right] 
+ \ln \ln \left[ \frac{4}{\pi} \left( \frac{\Delta}{2J} \right)^3 \right] \Bigg\} \\
+ O\left\{ \left[ \ln \left( \frac{\Delta}{2J} \right) \right]^{-1} \right\}.
\end{multline}
Thus, we find that the crossover time diverges logarithmically with $\Delta/J$ and, in the limit $\Delta/J \to \infty$, the crossover never occurs.
Indeed, the limit $\Delta/J \to \infty$ (which essentially makes the smallest gap $E(0) \to 0$) takes us to the original LZ model~\cite{Landau1932,Zener1932,Stuckelberg1932,Majorana1932}, and it implies the validity of the LZF for any value of $\tau$.
For any finite value $\Delta/J$, we can expect a transition to power-law decay for large enough $\tau$.

\section{\label{sec:IsingChain}Transverse-Field Ising chain}

Having demonstrated the application of APT to the simplest model, we now turn to the transverse-field Ising (TI) chain~\cite{Pfeuty1970}, a one-dimensional chain of $N$ spins.
This system possesses a QCP in the thermodynamic limit $N \to \infty$, where the gap between the ground and the first excited state vanishes.
Its Hamiltonian is
\begin{equation} \label{eq:TIHamiltonian}
H_\mathrm{TI}(\lambda) = - \frac{1}{2} \left( J \sum_{j=1}^{N} \sigma^z_j \sigma^z_{j+1} + \Gamma(\lambda) \sum_{j=1}^N \sigma^x_j \right),
\end{equation}
where the first sum represents the spin-spin interaction with coupling strength $J$ and the second sum represents the interaction of each spin to the external magnetic field $\Gamma(\lambda)$, rewritten for later convenience as
\begin{equation} \label{eq:TIExternalField}
\Gamma(\lambda) = J + \Delta \, \lambda.
\end{equation}
We assume periodic boundary conditions on the spins, $\sigma^z_{N+1} = \sigma^z_1$.

The Hamiltonian~\eqref{eq:TIHamiltonian} can be diagonalized exactly~\cite{Dziarmaga2005}.
For even $N$, and exploiting Jordan-Wigner, Fourier, and, finally, Bogoliubov transforms, we have~\cite{Dziarmaga2005}
\begin{equation} \label{eq:TIHamiltonianDiagonalized}
H_\mathrm{TI}(\lambda) = \sum_k \epsilon_k(\lambda) \left[ \gamma_k^\dag(\lambda) \gamma_k(\lambda) - 1/2 \right],
\end{equation}
where $\gamma_k^\dag(\lambda)$ and $\gamma_k(\lambda)$ are creation and annihilation operators of fermions with dispersion
\begin{equation} \label{eq:TIDispersion}
\epsilon_k(\lambda) = \sqrt{\left[ \Gamma(\lambda) - J\cos(ka) \right]^2 + J^2 \sin^2(ka)}. 
\end{equation}
The allowed values of $k$ are given by
\begin{equation} \label{eq:TIMomentumValues}
k_n =  \frac{\left( 2n + 1\right)\,\pi}{N a},
\end{equation}
where $n$ is an integer between $-N/2$ and $N/2 - 1$.
In the thermodynamic limit, $k$ is a continuous variable ranging from $-\pi/a$ to $\pi/a$, and sums can be replaced by integrals, $\sum_k \to N/2\pi\, \int_{-\pi}^{\pi} d(ka)$.

Equation~\eqref{eq:TIHamiltonianDiagonalized} describes free fermions with momentum $k$ and energy $\epsilon_k(\lambda)$.
Since the system is translationally invariant, total momentum must be conserved, and fermions can be created or destroyed only in pairs of opposite momenta $k$ and $-k$.
Therefore, if we start with the ground state (with no fermions), we restrict ourselves to only half of the momentum values.
In the limit $N \to \infty$, the lowest momentum $k_0 = \pi/N a \to 0$, and its energy vanishes when $\Gamma = J$ (or $\lambda = 0$), which signifies the QCP.

The ground state of the Ising chain can be expressed as
\begin{equation}
\ket{g(\lambda)} = \prod_{k>0} \left[ \cos\theta_k(\lambda) - \sin\theta_k(\lambda) c_k^\dag c_{-k}^\dag \right] \ket{\mbox{vac}},
\end{equation}
where $c_k \equiv \cos\theta_k \gamma_k - \sin\theta_k \gamma_{-k}^\dag$, $c_k \ket{\mbox{vac}} \equiv 0$ and
\begin{equation} \label{eq:TITheta}
\theta_k(\lambda) = \frac{1}{2} \arctan\left( \frac{J \sin(ka)}{\Gamma(\lambda) - J \cos(ka)} \right).
\end{equation}
Moreover, in complete analogy to the LZ model, a solution
\begin{equation} \label{eq:TIDynamicState}
\ket{\psi(t)} = \prod_{k>0} \left[ u_k(t) - v_k(t) c_k^\dag c_{-k}^\dag \right] \ket{\mbox{vac}}
\end{equation}
of the time-dependent Schr\"odinger equation can be determined from \cite{Dziarmaga2005}
\begin{equation} \label{eq:TIDifferentialEquations}
\begin{split}
i\, \dot{u}_k(t) &=  - \left[ \Gamma(\lambda) - J \cos(ka) \right] u_k(t) - J \sin(ka) v_k(t), \\
i\, \dot{v}_k(t) & =  - J \sin(ka) u_k(t) + \left[ \Gamma(\lambda) - J \cos(ka) \right] v_k(t).
\end{split}
\end{equation}

\begin{figure}
\includegraphics[width=\columnwidth]{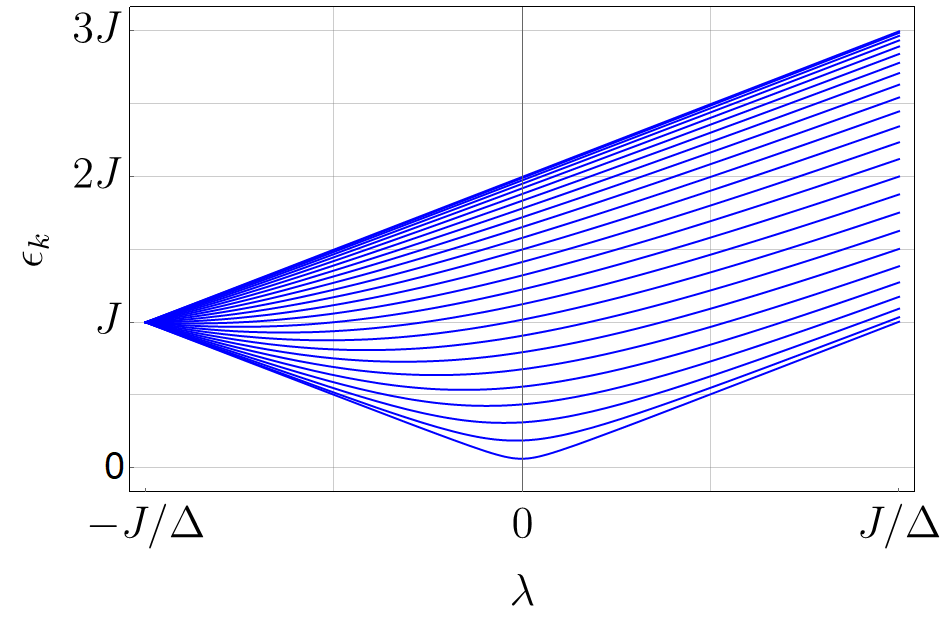}
\caption{\label{fig:TI_energysublevels}
Eigenenergies of the Ising chain in the transverse field \eqref{eq:TIDispersion} as a function of $\lambda$ for $N = 50$, where the lowest sublevel is given by $k_0 = \pi/Na$ and the highest is given by $k_{N/2-1}$.
Observe the avoided crossing around the critical point $\lambda = 0$ for the lowest-energy sublevels.
}
\end{figure}

Comparing Eqs.~\eqref{eq:LZDifferentialEquations} and \eqref{eq:TIDifferentialEquations}, we notice that the dynamics of the LZ and the TI model are formally identical if we identify
\begin{equation} \label{eq:TILZSubstitution}
\begin{split}
J & \Rightarrow  J_k = J \sin(ka), \\
\lambda & \Rightarrow  \lambda_k = \lambda + \frac{J}{\Delta} \left[ 1 - \cos(ka) \right].
\end{split}
\end{equation}
Therefore, the transverse-field Ising chain can be understood as $N/2$ generalized LZ models, where each two-level system corresponds to a (positive) value of $k$.
Consequently, when crossing the QCP, the sublevels that have $\Delta/J_k \gg 1$ go through an avoided crossing, as illustrated in Fig.~\ref{fig:TI_energysublevels}.

As before, we now compute the excess work~\eqref{eq:ExcessWork}.
We have
\begin{equation} \label{eq:TIExcessWork}
W_\mathrm{ex}(\tau) = \sum_{k>0} 2 \epsilon_k(\lambda_f) p_k(\tau),
\end{equation}
where
\begin{equation} \label{eq:TIExcitationProbability}
p_k(\tau) = \Big| \sin\theta_k(\lambda_f) u_k(t_f) - \cos\theta_k(\lambda_f) v_k(t_f) \Big|^2
\end{equation}
is the probability of creating a pair of fermions with opposite momenta $k$ and $-k$ during the evolution.
Again, for simplicity, we consider the linear protocol,
\begin{equation} \label{eq:TILinearProtocol}
\lambda(t) = \frac{t}{\tau}, \qquad -\frac{\tau}{2} \leq t \leq \frac{\tau}{2}.
\end{equation}

\begin{figure*}
\subfloat[\label{fig:TIWorkMainPlot}]{\includegraphics[width=.33\textwidth]{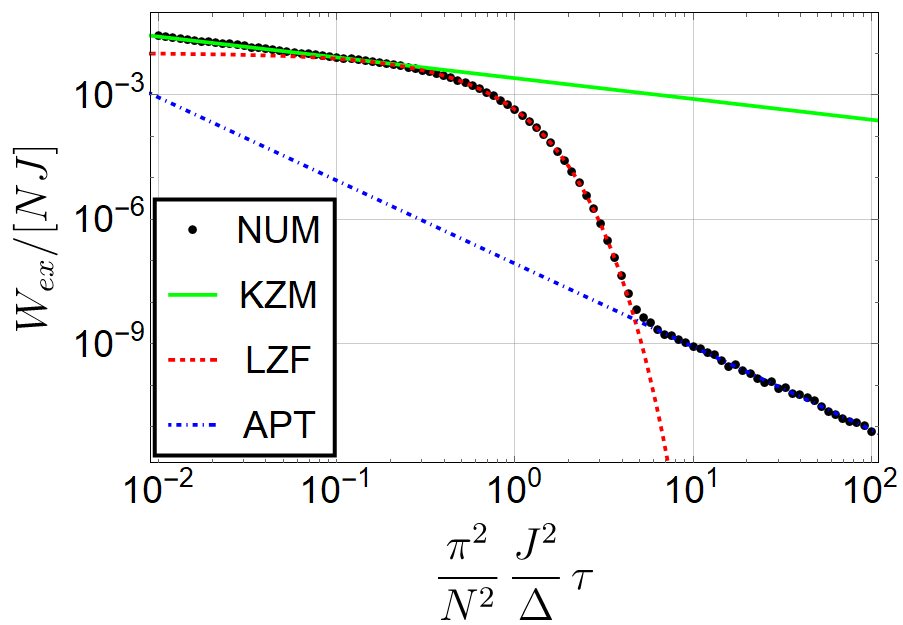}}
\subfloat[\label{fig:TIAPTOscillations}]{\includegraphics[width=.33\textwidth]{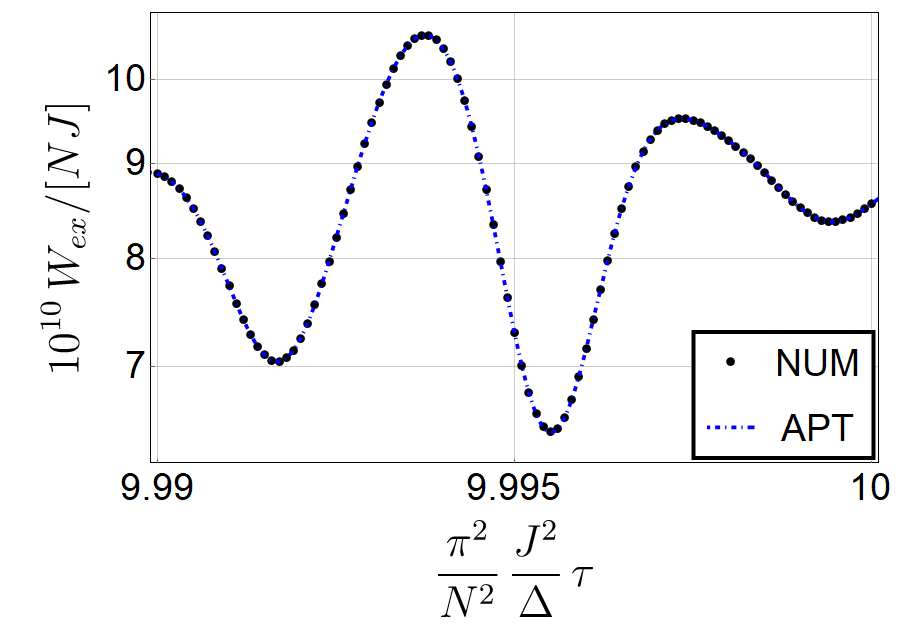}}
\subfloat[\label{fig:TIKZMFail}]{\includegraphics[width=.33\textwidth]{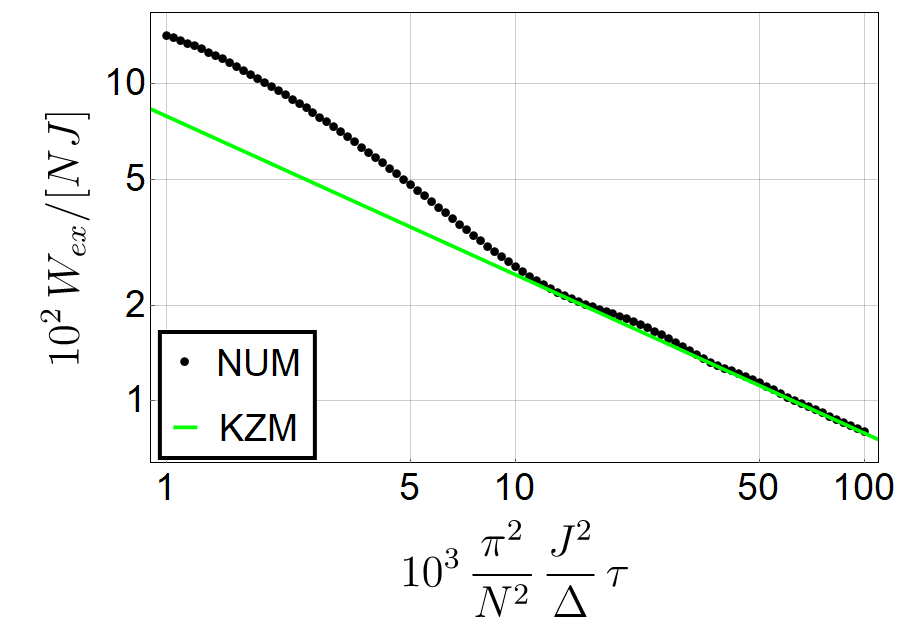}}
\caption{\label{fig:TIPlots}
Excess work~\eqref{eq:TIExcessWork} as a function of process duration for $N = 100$ and $\Delta/J = 1$.
Black dots represent the numerics, the green solid line represents Eq.~\eqref{eq:TI_KZMExcessWork}, the red dashed line represents Eq.~\eqref{eq:TI_LZFExcessWork}, and the blue dash-dotted line represents Eqs.~\eqref{eq:TI_APTExcessWork_continuous} and \eqref{eq:TI_APTExcessWork_discrete}.
(a) The two crossovers, KZM-LZF and LZF-APT.
(b) Zoom of a $\tau$ range where APT is valid, with oscillations included.
(c) Zoom of a $\tau$ range where KZM fails.
}  
\end{figure*}

The corresponding LZF \eqref{eq:LZFormula} becomes
\begin{equation} \label{eq:TILandauZenerFormula}
p_k^\mathrm{LZ}(\tau) = \exp\left[ -\pi J^2 \sin^2(ka) \tau/\Delta\right],
\end{equation}
where we exploited Eq.~\eqref{eq:TILZSubstitution}.
Note that this is valid for only the lowest-energy sublevels, which exhibit avoided crossings.
For these levels and in the limit $J^2 \tau/\Delta \gg 1$, we can employ small argument approximations in Eqs.~\eqref{eq:TIDispersion} and~\eqref{eq:TILandauZenerFormula}.
Thus, for $N\gg1$ the excess work~\eqref{eq:TIExcessWork} becomes
\begin{equation}
\begin{split}
\label{eq:wex_excat}
W_\mathrm{ex}(\tau) &= \frac{N}{\pi} \int_{0}^{\infty} \sqrt{\left( \Gamma_f - J \right)^2 + J \Gamma_f (ka)^2} \\
&\qquad\times \exp\left[ -\pi J^2 (ka)^2 \tau/\Delta \right] d(ka),
\end{split}
\end{equation}
where $\Gamma_f = \Gamma(\lambda_f)$ is the final value of the external field.
Equation~\eqref{eq:wex_excat} can be solved exactly in terms of hypergeometric functions.

In the limit $J^2 \tau/\Delta \gg 1$ an approximate expression reads
\begin{equation} \label{eq:TI_KZMExcessWork}
W_\mathrm{ex}^\mathrm{KZM}(\tau) = \frac{N\Delta|\lambda_f|}{2\pi}\,  \sqrt{\frac{\Delta}{J^2 \tau}}.
\end{equation}
The superscript KZM denotes the Kibble-Zurek mechanism \cite{Dziarmaga2005,Zurek2005,Polkovnikov2005}.
It has been shown that when crossing the QCP, arguments from the KZM \cite{Francuz2016PRB,Esposito2020} allow expressing the excess work in terms of the average number of excitations $n_\mathrm{ex}$,
\begin{equation} \label{eq:TI_KZMWork&Excitations}
W_\mathrm{ex}^\mathrm{KZM}(\tau) = 2 \Delta |\lambda_f| n_\mathrm{ex}.
\end{equation}

Equation~\eqref{eq:TI_KZMExcessWork} is valid if Eq.~\eqref{eq:TILandauZenerFormula} holds for the lowest sublevels and for $J^2 \tau/\Delta \gg 1$.
However, as $\tau$ increases, we reach a point $\frac{J^2 \tau}{\Delta} \left( \frac{\pi}{N} \right)^2 \sim 1$ where $p_k$~\eqref{eq:TILandauZenerFormula} is so highly peaked at $k_0$ that no other sublevel contributes to the sum in Eq.~\eqref{eq:TIExcessWork}.
In other words, Eq.~\eqref{eq:TILandauZenerFormula} holds only for $k_0$, and $p_k = 0$ for any other sublevel.
In this case, Eq.~\eqref{eq:TIExcessWork} becomes
\begin{equation} \label{eq:TI_LZFExcessWork}
W_\mathrm{ex}^\mathrm{LZF}(\tau) = 2 \Delta |\lambda_f| \exp\left[ -\pi \left( \frac{\pi}{N} \right)^2 \frac{J^2}{\Delta} \tau \right].
\end{equation}
Thus, we expect a crossover from the power-law decay of Eq.~\eqref{eq:TI_KZMExcessWork} to the exponential decay of Eq.~\eqref{eq:TI_LZFExcessWork}.

Finally, for even larger process duration $\tau$ we enter the range of validity of APT; namely, when $\frac{J^2 \tau}{\Delta} \left( \frac{\pi}{N} \right)^2 \gg 1$, APT must hold.
In this case, mirroring Eq.~\eqref{eq:LZ_APTTransitionProbability_Linear},
\begin{equation} \label{eq:TI_APTTransitionProbability_Linear}
p_k^\mathrm{APT}(\tau) = \frac{1}{16} \left( \frac{\Delta}{J_k^2\tau} \right)^2 \left| \frac{J_k^3}{\epsilon_k^3(\lambda_f)} -  \frac{J_k^3 \exp[-2i\phi_k(\tau)]}{\epsilon_k^3(\lambda_i)} \right|^2,
\end{equation}
where again $\phi_k(\tau) = - \tau \int_{\lambda_i}^{\lambda_f} \epsilon_k(\lambda) d\lambda$.

In this limit, the excess work~\eqref{eq:wex_excat} reads
\begin{equation} \label{eq:TI_APTExcessWork_discrete}
W_\mathrm{ex}^\mathrm{APT}(\tau) = \sum_{k>0} 2 \epsilon_k(\lambda_f) p_k^\mathrm{APT}(\tau).
\end{equation}
As for the LZ model, the dynamical phase $\phi_k(\tau)$ leads to a rapidly oscillating quantity.
However, for long spin chains, $N\gg1$, these oscillations average out, and we can write
\begin{equation} \label{eq:TI_APTExcessWork_continuous}
W_\mathrm{ex}^\mathrm{APT}(\tau) = \frac{NJ}{16\pi} \left( \frac{\Delta}{J^2\tau} \right)^2 f\left( \frac{\Delta}{J} \right),
\end{equation}
where
\begin{equation}
f\left( \frac{\Delta}{J} \right) = J^5 \int_{0}^{\pi} \sin^2(ka) \left( \frac{1}{\epsilon_k^5(\lambda_f)} + \frac{\epsilon_k(\lambda_f)}{\epsilon_k^6(\lambda_i)} \right) d(ka)
\end{equation}
is a unitless function that depends only on $\Delta/J$ and that can be written as sums of elliptic integrals. 

In Fig.~\ref{fig:TIPlots} we compare Eqs.~\eqref{eq:TI_KZMExcessWork}, \eqref{eq:TI_LZFExcessWork}, and \eqref{eq:TI_APTExcessWork_continuous}, with the numerically exact solution for $N = 100$ and $\Delta/J = 1$.
As for the LZ model we notice a distinct crossover from the LZF~\eqref{eq:TI_LZFExcessWork} to APT~\eqref{eq:TI_APTExcessWork_continuous}.
In complete analogy to the LZ model, we also find the dynamical oscillations, depicted in Fig.~\ref{fig:TIAPTOscillations}.

On the $x$ axis, we have chosen $\frac{\pi^2}{N^2} \frac{J^2}{\Delta} \tau$, as this makes it easy to identify the adiabatic regime.
For long chains, $N \gg 1$, the prefactor multiplying $\tau$ becomes very small, and hence, APT is applicable only for very slow processes.

The main difference between the LZ model and the TI chain comes from the size of the systems: the TI chain presents a power-law decay for $W_\mathrm{ex}$ for $\frac{\pi^2}{N^2} \frac{J^2}{\Delta} \tau < 1$, predicted by the KZM for $J^2\tau/\Delta  \gg 1$.
For even smaller values of $\tau$, it is known that this $\tau^{-1/2}$ scale breaks down because LZF ceases to be valid.
This can be seen in Fig.~\ref{fig:TIKZMFail}.
Note, however, that the value of $\tau$ where this breakdown happens decreases with increasing $N$.
This is because a larger $N$ makes the avoided crossing more pronounced, and as $N \to \infty$, the LZF is valid for any value $\tau$, as noted in the last paragraph of Sec.~\ref{sec:LandauZener}.

The condition $J^2 \tau/\Delta \gg 1$ for the validity of Eq.~\eqref{eq:TI_KZMExcessWork} is sometimes understood as a condition of adiabaticity since it requires large enough $\tau$.
We emphasize, however, that neither KZM~\eqref{eq:TI_KZMExcessWork} nor LZF \eqref{eq:TI_LZFExcessWork} is adiabatic in the strict sense.
Rather, the excess work exhibits two crossovers [see Fig.~\ref{fig:TIWorkMainPlot}]: from KZM ($W_\mathrm{ex} \sim \tau^{-1/2}$) to LZF [$W_\mathrm{ex} \sim \exp(-\alpha \tau)$]; and from LZF to APT ($W_\mathrm{ex} \sim \tau^{-2}$).
These crossovers have been identified and discussed many times before for the TI chain \cite{Dziarmaga2005,Morita2008,Bartek2018} and other systems \cite{Suzuki2005,Schaller2006,Rezakhani2010,Wauters2017,Passarelli2018,Xue2018,Srivastava2020}.

The main contribution of our detailed analysis is the quantification of the crossover points.
The KZM-LZF crossover time $\tau_{1}$ and LZF-APT crossover time $\tau_{2}$ can be estimated in complete analogy to that discussed above.
For KZM-LZF, we equate Eqs.~\eqref{eq:TI_KZMExcessWork} and~\eqref{eq:TI_LZFExcessWork}.
Solving for $\tau_{1}$ results in a complex value since the two curves never intersect [see Fig.~\ref{fig:TIWorkMainPlot}].
For the real part, we obtain
\begin{equation} \label{eq:TICrossoverTime1}
\frac{\pi^2}{N^2} \frac{J^2}{\Delta} \tau_{1} = - \frac{1}{2\pi} \re{ W_{-1} \left( - \frac{\pi}{8} \right) } \approx 0.152,
\end{equation}
where $W_{-1}$ is again Lambert's function~\cite{Lambert1996}.
Consequently, the location of the KZM-LZF crossover is independent of $N$ in Fig.~\ref{fig:TIPlots}.

For LZF-APT, we equate Eqs.~\eqref{eq:TI_LZFExcessWork} and \eqref{eq:TI_APTExcessWork_continuous}.
We obtain, for $\tau_{2}$ in the limit of $N\gg1$,
\begin{equation} \label{eq:TICrossoverTime2}
\begin{split}
\frac{\pi^2}{N^2} \frac{J^2}{\Delta} \tau_{2} &= \frac{2}{\pi} \Bigg\{ \ln \left( \frac{4}{\pi} \left[ \frac{J}{4\Delta} f\left( \frac{\Delta}{J} \right) \right]^{-1/2} \left[ \frac{N}{\pi} \right]^{3/2} \right) \\
&+ \ln \ln \left( \frac{4}{\pi} \left[ \frac{J}{4\Delta} f\left( \frac{\Delta}{J} \right) \right]^{-1/2} \left[ \frac{N}{\pi} \right]^{3/2} \right) \Bigg\} \\
&+ O\left[ \left( \ln N \right)^{-1} \right],
\end{split}
\end{equation}
Equation~\eqref{eq:TICrossoverTime2} is similar in form to Eq.~\eqref{eq:LZCrossoverTime}.
For $N \to \infty$, the crossover time diverges, which is the same as saying that the crossover never happens.
This is consistent with the fact that, when the gap vanishes, no evolution can be adiabatic and, therefore, APT always fails.
However, for any finite $N$, adiabaticity and power-law scaling $\tau^{-2}$ can always be attained for large enough $\tau$.

\paragraph*{Linear response theory.}

\begin{figure}
\includegraphics[width=\columnwidth]{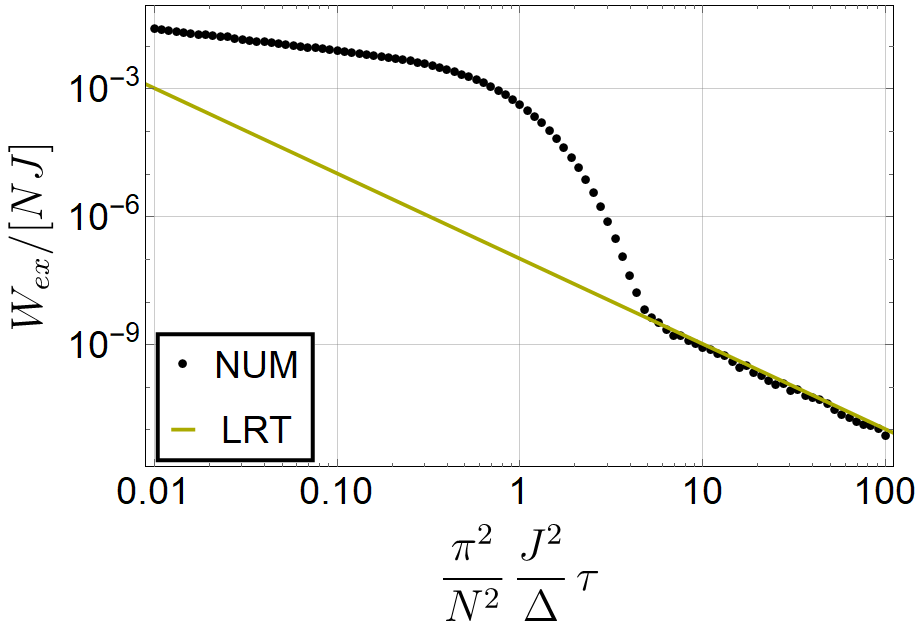}
\caption{\label{fig:TIPlot_LRT}
Excess work \eqref{eq:TIExcessWork} as a function of process duration for $N = 100$ and $\Delta/J = 1$.
Black dots represent the numerics, and the dark yellow line represents the result from LRT~\eqref{eq:ExcessWorkLRT}.
}
\end{figure}

We conclude this section by highlighting that the $\tau^{-2}$ scaling, derived from APT, can also be obtained using a linear response theory (LRT) approach~\cite{Kubo1985}.
In this framework, the excess work is expressed as (see Appendix~\ref{app:ExcessWorkLRT} for more details)
\begin{equation}
    W_{\rm ex}^\mathrm{LRT}(\tau) = \Delta^2 \int_{t_i}^{t_f}\int_{t_i}^t\Psi_0(t-t')\dot{\lambda}(t)\dot{\lambda}(t')dtdt',
\end{equation}
where the relaxation function $\Psi_0(t)$ is obtained from the response function $\Phi_0(t)$,
\begin{equation}
    \Phi_0(t) = -i\langle[\partial_{\Gamma}H(0),\partial_{\Gamma}H(t)]\rangle,
    \label{eq:rspfunc}
\end{equation}
using the relation $\Phi_0(t) = -d\Psi_0(t)/dt$~\cite{Kubo1985} (the symbol $[\cdot,\cdot]$ denotes the commutator).
We remark that the time evolution in Eq.~\eqref{eq:rspfunc} is obtained from the solutions of Heisenberg's equations with the initial Hamiltonian.

Using the transformations of Ref.~\cite{Dziarmaga2005} mentioned in Sec.~\ref{sec:IsingChain}, we can show that~\cite{Naze2021}
\begin{equation}
    \Psi_0(t) = \sum_{k>0}\frac{ J^2}{\epsilon^3_k(\lambda_i)}\sin^2{\left(k a\right)}\cos{\left[2\epsilon_k(\lambda_i) t \right]},
\end{equation}
Therefore, the excess work is
\begin{equation}
    W_\mathrm{ex}^\mathrm{LRT}(\tau)=\frac{J^2}{\tau^2}\left(\frac{\Delta}{2}\right)^2\sum_{k>0}\frac{1-\cos{[2\epsilon_k(\lambda_i) \tau]}}{\epsilon_k^5(\lambda_i)}\sin{\left(k a\right)},
    \label{eq:ExcessWorkLRT}
\end{equation}
which scales like $\tau^{-2}$ for large switching times.
Figure~\ref{fig:TIPlot_LRT} compares the numerical results with those of LRT, where we have again suppressed the dynamical oscillations for ease of presentation.
We notice that LRT provides the correct scale, although with a small shift from the exact values, which speaks to the validity of LRT for the specific values of $\Delta$ and $J$ used~\cite{Naze2021}.
Despite the reasonable performance of LRT for large $\tau$, APT is better fitted to calculate the crossover time from LZF, and it is easier to generalize in the case of nonlinear, two-parameter protocols, which are frequently encountered in realistic settings (see the next section).

\section{\label{sec:DWave}Quantum annealing}

\begin{figure}

\includegraphics[width=\columnwidth]{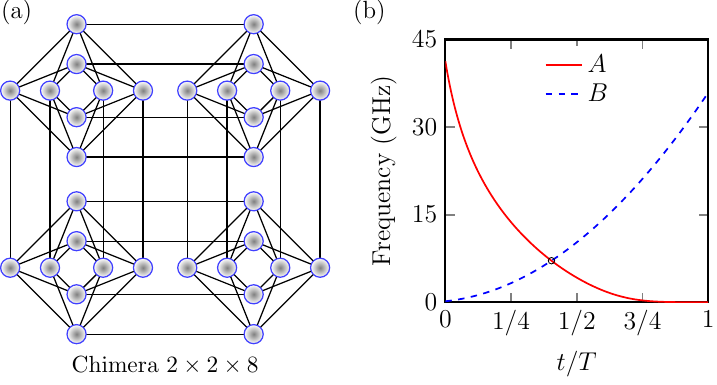}

\caption{\label{fig:dwave}
D-Wave annealing processor specification.
(a) Sparse Chimera graph (denoted as $C_2$), consisting of a $2 \times 2$ grid of clusters (i.e., unit cells) of eight qubits each.
The maximum number of qubits for this topology is $N=2048$ ($C_{16}$), whereas the number of all connections between them is $6000 \ll N^2$.
(b) A typical annealing schedule, where $T$ denotes the time to complete one annealing cycle.
}

\end{figure}

In the previous section, we discussed how to determine the crossover times in the TI chain.
While these can be dismissed for large systems, recent developments in the manipulation of small quantum systems make the crossovers achievable.
For instance, the D-Wave $2000Q$ (and later) quantum annealers~\cite{Bartek2018} realize the following time-dependent transverse-field Ising Hamiltonian:
\begin{equation}
\label{eq:dwave}
\mathcal{H}(t)/(2\pi\hbar)= -A(t) \mathcal{H}_0  -B(t) \mathcal{H}_{\text{Ising}},
\quad
t \in [0, T],
\end{equation}
where its classical part $\mathcal{H}_{\text{Ising}}$ is defined on a particular graph $\mathcal{G} = (\mathcal{V}, \mathcal{E})$ specified by its edges and vertices [see Fig.~\ref{fig:dwave}(a), where the Chimera graph $C_2$ is shown], as
\begin{equation}
  \label{eq:Hproblem}
  \mathcal{H}_{\text{Ising}} : = 
  \sum_{\langle i, j\rangle \in \mathcal{E}} J_{ij} \sigma_i^z \sigma_j^z + \sum_{i\in\mathcal{V}}h_i \sigma_i^z,
\end{equation}
whereas the ``free'' part reads
\begin{equation}
    \mathcal{H}_0 = \sum_{i \in \mathcal{V}} \sigma_i^x.
\end{equation}
The programmable annealing time $T$ varies from microseconds ($\sim2\,\mu$s) to milliseconds ($\sim2000\,\mu$s) depending on the specific schedule, which can also vary between devices~\cite{Bartek2018}.
A typical annealing schedule is shown in Fig.~\ref{fig:dwave}(b).
During the experiment, $A(t)$ changes from $A(0) \gg 0$ (i.e., all spins point in the $x$ direction) to $A(T)\approx 0$, whereas $B(t)$ is varied from $B(0)\approx 0$ to $B(T) \gg 0$ (i.e., $\mathcal{H}(T) \sim \mathcal{H}_{\text{Ising}}$).
Defining a one-dimensional path on the graph $\mathcal{G}$, putting $h_i=0$ and $J_{ij}=0$ for all spins not in that path, we can realize the TI model with two time-dependent parameters.

Therefore, the D-Wave setup supports a considerable range of number of spins $N$ (up to $|\mathcal{V}| \sim 5000$ with $|\mathcal{E}| \sim 40\,000$ for the Pegasus topology~\cite{dattani.szalay.19}), and process durations $\tau$ to test the crossover times of Eqs.~\eqref{eq:TICrossoverTime1} and \eqref{eq:TICrossoverTime2}.
The excess work of the annealing protocol is then calculated from the final energy, which can be read directly from the D-Wave solver.

\begin{figure}
\includegraphics[width=\columnwidth]{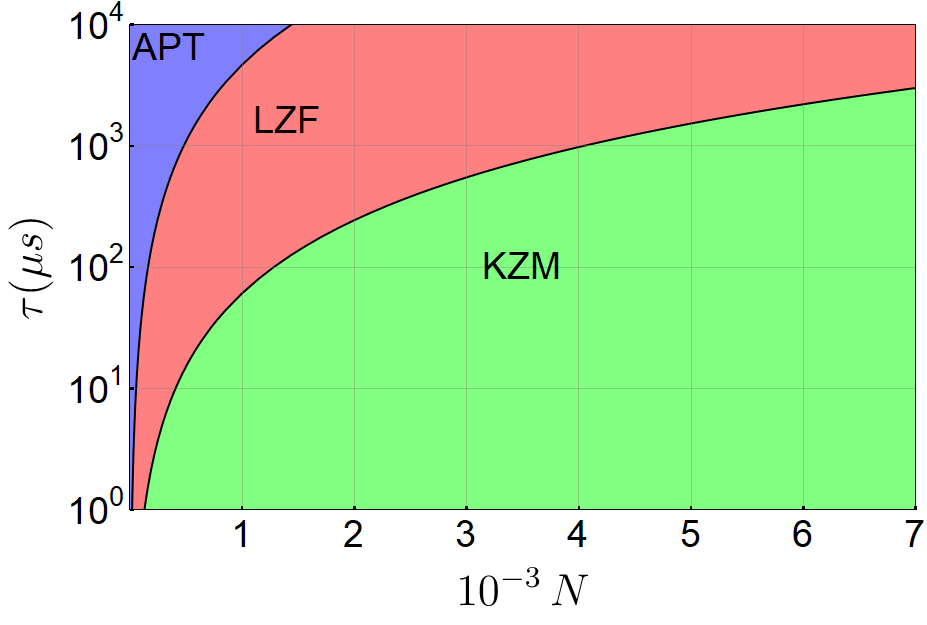}
\caption{\label{fig:TIScaleDiagram}
Phase diagram of scaling behaviors when crossing the QCP of the TI chain with realistic D-Wave parameters, as demonstrated in Fig.~\ref{fig:dwave}(b).
The values of $\tau$ and $N$ can be tuned in a given realization of the process.
The solid curves represent the crossover points between KZM, LZF, and APT.
}
\end{figure}

Figure~\ref{fig:TIScaleDiagram} is a corresponding ``phase'' diagram of the TI chain.
If the pair $(N,\tau)$ lies in the green region (KZM), the excess work behaves as $\tau^{-1/2}$.
If it lies in the red region (LZF), the excess work decays exponentially with $\tau$.
And if it lies in the blue region (APT), the excess work scales like $\tau^{-2}$.

Thus, our theoretical prediction can be experimentally verified on the D-Wave machine.
Equations~\eqref{eq:TICrossoverTime1} and~\eqref{eq:TICrossoverTime2} can be generalized for the case of two time-dependent parameters.
The number of spins would have to be kept low to have feasible times greater than $\tau_1$ of Eq.~\eqref{eq:TICrossoverTime1} and $\tau_2$ of Eq.~\eqref{eq:TICrossoverTime2}, but not so low that the lowest-energy sublevel does not go through an avoided crossing.
Once $N$ is decided, diagrams like that in Fig.~\ref{fig:TIScaleDiagram} would then provide the $\tau$ range to explore on D-Wave.
For example, with $500$ spins the KZM-LZF crossover would be around $10\,\mu$s, and the LZF-APT crossover would be around $10^3 \,\mu$s.

Finally, we remark that, while we offered here an analysis of the implications for D-Wave, the discussed phenomena should be verifiable in any quantum simulator that can implement the TI chain, as long as it can emulate the adiabatic process itself (see Ref.~\cite{Rams2021} for an exception).
It also should be noted that in any realistic quantum annealer one will have to contend with effects of environmental noise.
For instance, Ref.~\cite{King2022} reported for similar sized chains and in the weak coupling regime a coherence time of $10^{-1}\,\mu$s, after which excitations from the environment are significant and the dynamics can no longer be considered unitary.
Thus, the coherence time is much shorter than the driving times at which we predict the crossovers.
However, powerful quantum-error-correcting schemes exist \cite{Sarovar2013,Pudenz2015,Pastawski2016,Jordan2006,Jiang2017,Marvian2017}, even if some of them are still out of reach for currently available hardware.
For an experimental exploration of the here-predicted crossover behavior the implementation of viable error-correction schemes may be necessary to be able to cleanly distinguish between diabatic excitations and thermal noise.

\section{\label{sec:Conclusion}Concluding remarks}

It has been argued~\cite{Sanders2017} that all viable architectures for quantum computing will necessitate the implementation of quantum-error-correcting codes~\cite{nielsen2010}.
For quantum annealers this poses special challenges, as they experience two fundamentally different sources of error~\cite{Sarandy2005PRL,Young2013}: environmental noise and nonadiabatic excitations.
Whereas effective algorithms to mitigate the effects of environmental noise already exist~\cite{Sarovar2013,Pudenz2015,Pastawski2016,Vinci2018}, circumventing the consequences of finite-time driving is a much harder task.
In principle, so-called shortcuts to adiabaticity~\cite{Chen2010,Campo2012PRL,Santos2015SR,Acconcia2015PRE,Guery2019,carolan2021,Touil2021entropy} may hold the solution, but typically, the required control fields are highly complex and nonlocal. 

Thus, a comprehensive characterization of the finite-time excitations is instrumental if the ``outcome'' of a computation on a quantum annealer is to be trusted.
For the one-dimensional Ising chain in the transverse field, this is exactly what we have achieved in Fig.~\ref{fig:TIScaleDiagram}.
We verified and quantified earlier findings that indicated crossovers from a regime of the Kibble-Zurek mechanism to effective Landau-Zener dynamics to a third regime fully described by adiabatic perturbation theory.
This allowed us to unambiguously determine the crossover points, that is, the driving times $\tau$, for which the scaling properties of the excess work fundamentally change.
Thus, we expect our results to be directly and immediately applicable in the characterization of all present and future quantum annealers.

\begin{acknowledgments}
A.S. acknowledges support from the National Council for Scientific and Technological Development (CNPq, Brazil) under Grant No. 140549/2018-8.
P.N. and M.V.S.B. acknowledge financial support from CNPq under Grant No. 141018/2017-8 and FAPESP (Funda\c{c}\~ao de Amparo \`a Pesquisa do Estado de S\~ao Paulo, Brazil; Grants No. 2018/06365-4, No. 2018/21285-7, and No. 2020/02170-4).
B.G. acknowledges support from the National Science Center (NCN), Poland, under Projects No. 2020/38/E/ST3/00269 and the PL-Grid infrastructure.
S.D. acknowledges support from the U.S. National Science Foundation under Grant No. DMR-2010127.
\end{acknowledgments} 

\appendix

\section{\label{sec:HalfCrossing}Stopping at the QCP}

In the above analysis, we focused on driving protocols that are symmetric with respect to the quantum critical point.
In this appendix, we briefly outline the ramifications of stopping right at the QCP.
To this end, we consider the protocol
\begin{equation} \label{eq:HLZLinearProtocol}
\lambda(t) = \frac{t}{\tau}, \qquad -\tau \leq t \leq 0\,.
\end{equation}

\paragraph*{Landau-Zener model.}

As before, we first analyze the LZ model with Hamiltonian \eqref{eq:LZHamiltonian}.
Curiously, these situations are more involved, as there is no simple formula for the transition probability.
Rather, we have a ``half'' LZ formula (HLZ), which is given by a rather complicated expression \cite{Damski2006},
\begin{equation} \label{eq:HLZFormula}
\begin{split}
&p_+^\mathrm{HLZ}(\tau) = 1 -  \frac{\sinh\left( \pi\, J^2 \tau/2\Delta \right)}{\pi\,J^2 \tau/\Delta }\,\exp\left( -\pi\,J^2 \tau/4\Delta \right) \\
&\times \left| \Gamma\left( 1 + \frac{i}{4} \frac{J^2 \tau}{\Delta} \right) + \frac{e^{i \pi/4}}{2} \sqrt{ \frac{J^2 \tau}{\Delta} } \Gamma\left( \frac{1}{2} + \frac{i}{4} \frac{J^2 \tau}{\Delta} \right)\right|^2.
\end{split}
\end{equation}
Here, $\Gamma$ represents the Gamma function.

Equation~\eqref{eq:HLZFormula} holds for an infinite-time protocol with a nonzero derivative~\cite{Damski2006}, and thus, it also applies to our protocol~\eqref{eq:HLZLinearProtocol} for $\Delta/J \gg 1$ and $\left( \frac{\Delta}{2J} \right)^2 \frac{J^2 \tau}{\Delta} \gg 1$.
However, in contrast to the symmetric case, the HLZ includes the APT limit, obeying $p_+ \sim \tau^{-2}$ for $J^2 \tau/\Delta \gg 1$.

\begin{figure*}
\subfloat[\label{fig:HLZWorkMainPlot}]{\includegraphics[width=\columnwidth]{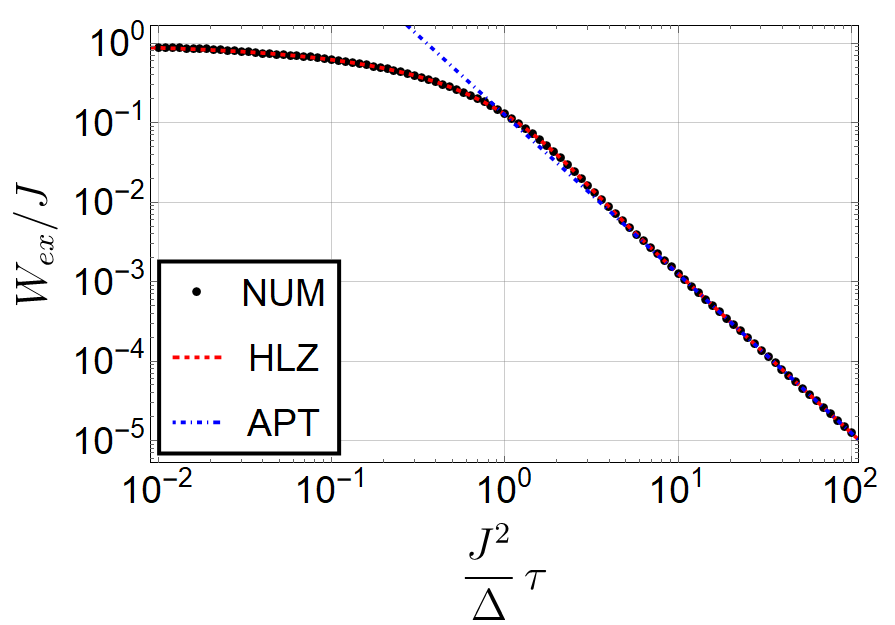}}
\subfloat[\label{fig:HLZ_HLZFail}]{\includegraphics[width=\columnwidth]{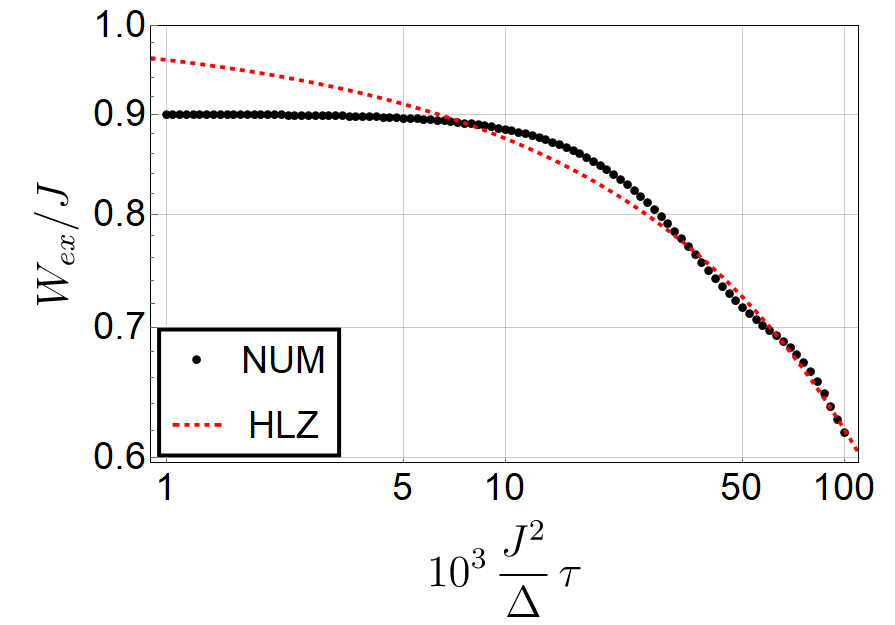}}
\caption{\label{fig:HLZPlot}
Excess work \eqref{eq:LZExcessWork} as a function of process duration for $\Delta/J = 10$ and for Eq.~\eqref{eq:HLZLinearProtocol}.
Black dots represent the numerics, the red dashed line represents the excess work calculated with the HLZ~\eqref{eq:HLZFormula}, and the blue dash-dotted line represents the excess work calculated from APT~\eqref{eq:LZ_APTTransitionProbability_Linear}.
(a) The agreement between HLZ and APT for large enough $\tau$.
(b) Zoom of a $\tau$ range where HLZ fails.
}
\end{figure*}

From the point of view of APT, the calculations for the excess work~\eqref{eq:ExcessWork} are the same as the case of crossing the $\lambda = 0$ point.
The transition probability is still given by Eq.~\eqref{eq:LZ_APTTransitionProbability_Linear}, but with different $\lambda_i$ and $\lambda_f$.
In Fig.~\ref{fig:HLZPlot} we compare $W_\mathrm{ex}$~\eqref{eq:LZExcessWork} with $p_+$ calculated in three ways: with HLZ, with APT, and with numerical evolution.
Figure~\ref{fig:HLZWorkMainPlot} demonstrates very good agreement of HLZ with the numerics for the entire range of the plot while also showing that it agrees with APT for large enough $\tau$.
The oscillations present in APT still exist, but they are tamer and, in this specific example, invisible.
In Fig.~\ref{fig:HLZ_HLZFail} we highlight that HLZ does, indeed, fail for small enough $\tau$.
Finally, since HLZ and APT agree, there is no crossover.

\paragraph*{Ising chain in the transverse field.}

Now, we turn our attention to the TI chain \eqref{eq:TIHamiltonianDiagonalized}, using the same protocol \eqref{eq:HLZLinearProtocol}.
We begin with the prediction from KZM.
To this end, we employ Eq.~\eqref{eq:TIExcessWork} with $\lambda_f = 0$ and $p_k(\tau)$ given by Eq.~\eqref{eq:HLZFormula} with the substitutions from Eq.~\eqref{eq:TILZSubstitution}.
This is valid only for the lowest-energy sublevels, which obey $\Delta/J_k \gg 1$.
Again, approximating sums by integrals, applying small argument approximations in the trigonometric functions, extending the upper integral limit to infinity, and defining a new variable of integration $x = \sqrt{J^2\tau/\Delta}\, k a$, Eq.~\eqref{eq:TIExcessWork} becomes
\begin{equation} \label{eq:HTI_KZMExcessWork}
W_\mathrm{ex}^\mathrm{KZM}(\tau) = \frac{K N J}{\pi} \frac{\Delta}{J^2 \tau},
\end{equation}
where
\begin{equation} \label{eq:HTI_KZMConstant}
\begin{split}
&K \equiv \int_0^{\infty} x \Bigg[ 1 - \exp\left( -\frac{\pi}{4} x^2 \right) \frac{\sinh\left( \frac{\pi}{2} x^2 \right)}{\pi x^2 } \\
&\times \left| \Gamma\left( 1 + \frac{i}{4} x^2 \right) + \frac{\exp(i \pi/4)}{2} x \Gamma\left( \frac{1}{2} + \frac{i}{4} x^2 \right)\right|^2 \Bigg] dx
\end{split}
\end{equation}
is an integral that can be computed numerically.

Note that, when stopping at the QCP, KZM dictates $W_\mathrm{ex} \sim \tau^{-1}$, which is different from the KZM result when crossing the QCP \cite{Francuz2016PRB,Esposito2020}.
In particular, $W_\mathrm{ex}^\mathrm{KZM}(\tau)$~\eqref{eq:HTI_KZMExcessWork} is \emph{not} proportional to the average number of excitations $n_\mathrm{ex}$, which scales like $\tau^{-1/2}$ for the present $\tau$ range.

On the other hand, the calculations from APT once again follow the expressions of crossing the QCP.
The excitation probability is given by Eq.~\eqref{eq:TI_APTTransitionProbability_Linear}, with $\lambda_i = -1$ and $\lambda_f = 0$, in accordance with Eq.~\eqref{eq:HLZLinearProtocol}.
Note, however, that the first term inside the absolute value diverges for $k = 0$.
This means that for $N\gg1$ the excess work is dominated by the lowest-energy sublevel, and we have
\begin{equation} \label{eq:HTI_APTExcessWork}
W_\mathrm{ex}^\mathrm{APT}(\tau) = \frac{N J}{8\pi} \left( \frac{N}{\pi} \right)^2 \left( \frac{\Delta}{J^2 \tau} \right)^2.
\end{equation}

\begin{figure*}
\subfloat[\label{fig:HTIWorkMainPlot}]{\includegraphics[width=\columnwidth]{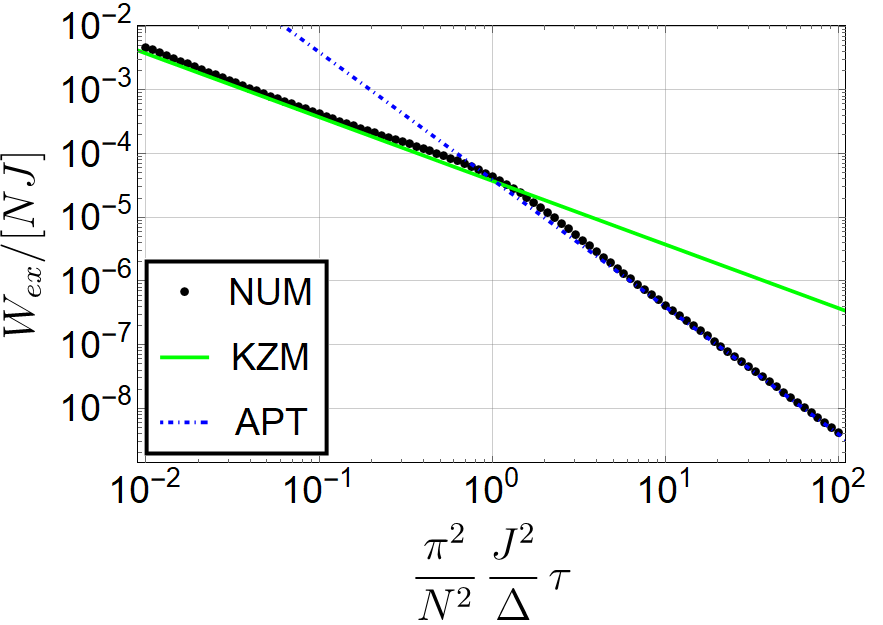}}
\subfloat[\label{fig:HTI_KZMFail}]{\includegraphics[width=\columnwidth]{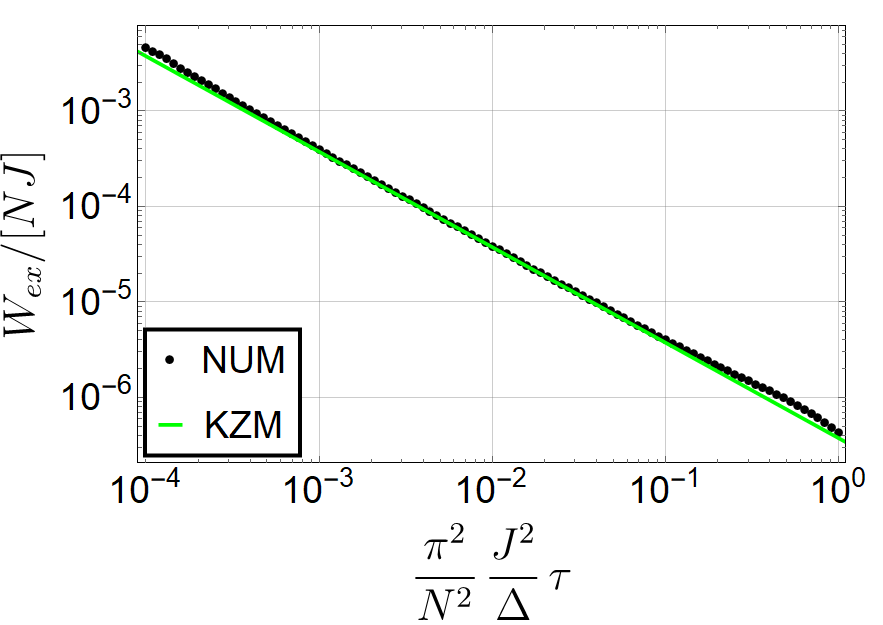}}
\caption{\label{fig:HTIPlots}
Excess work~\eqref{eq:TIExcessWork} as a function of process duration for $\Delta/J = 1$ and for Eq.~\eqref{eq:HLZLinearProtocol}.
Black dots represent the numerics, the green solid line represents Eq.~\eqref{eq:HTI_KZMExcessWork}, and the blue dash-dotted line represents the excess work calculated from APT \eqref{eq:HTI_APTExcessWork}.
(a) The crossover from KZM to APT for $N=100$.
(b) The KZM prediction compared to numerics for $N=1000$.
}
\end{figure*}

In Fig.~\ref{fig:HTIPlots} we show the resulting $W_\mathrm{ex}$ from KZM~\eqref{eq:HTI_KZMExcessWork} and from APT~\eqref{eq:HTI_APTExcessWork}, together with the numerically exact results.
Observe in Fig.~\ref{fig:HTIWorkMainPlot} that for $N=100$, the situation is similar to what we have discussed above: APT matches the numerical findings for $\frac{\pi^2}{N^2} \frac{J^2 \tau}{\Delta} \gg 1$, while KZM gives the correct behavior for $\frac{\pi^2}{N^2} \frac{J^2 \tau}{\Delta} \ll 1$.
The agreement between Eq.~\eqref{eq:HTI_KZMExcessWork} and numerics becomes even more convincing for larger systems, as demonstrated in Fig.~\ref{fig:HTI_KZMFail}.

In conclusion, we find a KZM-APT crossover when stopping at the QCP.
The crossover time $\tau_c$ can be estimated from Eqs.~\eqref{eq:HTI_KZMExcessWork} and \eqref{eq:HTI_APTExcessWork}, and we obtain
\begin{equation}
\frac{\pi^2}{N^2} \frac{J^2}{\Delta} \tau_c = \frac{1}{8K} \approx 1.049\,,
\end{equation}
which is again independent of $N$.

\section{Excess work from LRT}
\label{app:ExcessWorkLRT}

Finally, we briefly outline how to obtain the excess work~\eqref{eq:ExcessWork} from linear-response theory.
To this end, consider a quantum system that is in contact with a heat bath of temperature $\beta \equiv (k_B T)^{-1}$, where $k_B$ is Boltzmann's constant.
As before, during a switching time $\tau$, the external parameter is changed from $\lambda_i$ to $\lambda_i+\delta\lambda$.
The average work performed on the system during this process is \cite{jarzynski1997}
\begin{equation}
W \equiv \int_{t_i}^{t_f} \overline{\partial_{\lambda}H}(t)\dot{\lambda}(t)dt\,.
\label{eq:work}
\end{equation}
The generalized force $\overline{\partial_{\lambda}H}$ is calculated from the trace
\begin{equation}
\overline{\partial_{\lambda}H}(t) = \tr{\rho(t) \partial_{\lambda}H},
\end{equation}
where $\rho(t)$ is a nonequilibrium density matrix evolved under the von Neumann--Liouville equation.
The external parameter can be expressed as
\begin{equation}
\lambda(t) = \lambda_0+h(t)\delta\lambda,
\end{equation}
where the protocol $h(t)$ must satisfy the following boundary conditions:
\begin{equation}
h(t_i)=0,\quad h(t_f)=1.
\label{eq:bc}
\end{equation}

Linear-response theory aims to express average quantities to first order in the perturbation parameter $\delta\lambda/\lambda_{0}$ considering how this perturbation affects the observable to be averaged and the nonequilibrium state $\rho(t)$.
In our case, we assume that the parameter does not change considerably during the process, i.e., $|h(t)\delta\lambda/\lambda_0|\ll 1$ for all $t\in[t_i,t_f]$.
Thus, the generalized force can be expressed as \cite{Kubo1985}
\begin{equation}
\begin{split}
\overline{\partial_{\lambda}H}(t)& = \langle\partial_{\lambda}H\rangle_0+\delta\lambda\langle\partial_{\lambda}^2 H\rangle_0 h(t) \\
&-\delta\lambda\int_{t_i}^t \Phi_0(t-t')h(t')dt',
\end{split}
\label{eq:genforce-resp}
\end{equation}
where $\langle\cdot\rangle_0$ is the average over the initial canonical ensemble.
The quantity $\Phi_0(t)$ is the so-called response function~\cite{Kubo1985}, which can be conveniently expressed as the derivative of the relaxation function $\Psi_0(t)$,
\begin{equation}
\Phi_0(t) = -\dot{\Psi}_0(t).
\end{equation} 

The generalized force, written in terms of the relaxation function, reads
\begin{equation}
\begin{split}
\overline{\partial_{\lambda}H}(t) &= \langle\partial_{\lambda}H\rangle_0-\delta\lambda\widetilde{\Psi}_0 h(t) \\
&+\delta\lambda\int_{t_i}^t \Psi_0(t-t')\dot{h}(t')dt',
\end{split}
\label{eq:genforce-relax}
\end{equation}
where $\widetilde{\Psi}_0\equiv \Psi_0(0)-\langle\partial_{\lambda\lambda}^2H\rangle_0$.
Finally, combining Eqs.~\eqref{eq:work} and~\eqref{eq:genforce-relax}, the average work becomes
\begin{equation}
\begin{split}
W^\mathrm{LRT}& = \delta\lambda\langle\partial_{\lambda}H\rangle_0-\frac{\delta\lambda^2}{2}\widetilde{\Psi}_0\\
&+\delta\lambda^2 \int_{t_i}^{t_f}\int_{t_i}^t \Psi_0(t-t')\dot{h}(t')\dot{h}(t)dt'dt.
\label{eq:work2}
\end{split}
\end{equation}

It can be shown that the first two terms of Eq.~\eqref{eq:work2} [those independent of the protocol $h(t)$] give exactly the quasistatic work, i.e., the work performed if the process were quasistatic, when $\delta\lambda/\lambda_0 \ll 1$~\cite{bonanca2015}.
Thus, we define 
\begin{equation}
\begin{split}
W_\mathrm{ex}^\mathrm{LRT} = \delta\lambda^2 \int_{t_i}^{t_f}\int_{t_i}^t \Psi_0(t-t')\dot{h}(t')\dot{h}(t)dt'dt,
\label{eq:wirrftwp}
\end{split}
\end{equation}
as the LRT expression for the excess work.
This is the expression used in Eq.~\eqref{eq:ExcessWorkLRT}.

\bibliography{bibliography}
\bibliographystyle{apsrev4-2}

\end{document}